\documentclass[noshowpacs,preprint,aps,pre]{revtex4}
\usepackage[dvips]{graphicx}
\usepackage{latexsym}
\usepackage{epsfig}
\usepackage{bm}
\usepackage{color}     
\usepackage{times,psfrag,subfigure}        
\begin{document}

\title{Equilibrium morphologies and effective spring constants \\ of 
capillary bridges}
\author{H. Kusumaatmaja}
\author{R. Lipowsky}
\affiliation{Department of Theory and Bio-Systems, Max Planck Institute of Colloids and Interfaces, 14424 Potsdam, Germany}
\date{\today}
%%%%%%%%%%%%%%%%%%%%%%%%%%%%%%%%%%%%%%%%%%%%%%%%%%%%%%%%%%%%%%%%%%%%%%%%%%%%%

\begin{abstract}

We theoretically study the behaviour of a liquid bridge formed between a pair of rigid and parallel plates. The plates are smooth, they may either be homogeneous or decorated by circular patches of more hydrophilic domains, and they are generally not identical. We calculate the mechanical equilibrium distance of the liquid bridge as a function of liquid volume, contact angle and radius of the chemical domain. We show that a liquid bridge can be an equilibrium configuration as long as the sum of the contact angles at the two walls is larger than $180^\circ$. When comparisons are possible, our results agree well with recent analytical and molecular dynamics simulation results. We also derive the effective spring constant of the liquid bridge as it is perturbed from its equilibrium distance. The spring constant diverges when the sum of the contact angles is $180^\circ$ and is finite otherwise. The value of the spring constant decreases with increasing contact angle and volume, and the rate, at which it decreases, depends strongly on the properties of the two plates.

\end{abstract}
%\pacs{68.08.Bc}{Wetting}
%\pacs{68.03.Cd}{Surface tension and related phenomena}
%\pacs{68.35.Ct}{Interface structure and roughness}
\maketitle

%%%%%%%%%%%%%%%%%%%%%%%%%%%%%%%%%%%%%%%%%%%%%%%%%%%%%%%%%%%%%%%%%%%%%%%%%%%%%%%%%%%%%%%%%%%%%%%%%%%%%%%%%

\section{Introduction}

Recently there has been a resurgence of interest in capillary phenomena. One area that has come to the fore because of advances in patterning techniques is the behaviour of liquids in contact with structured surfaces. These wetting systems are extremely interesting because (i) they often show rich static and dynamical behaviours, such as non-spherical drop shapes \cite{Brinkmann}, morphological wetting transitions \cite{Lenz,Seemann}, and hysteretic phenomena \cite{Kusumaatmaja, Johnson}, and (ii) further understanding in these phenomena leads to smart designs of technological applications. The latter is particularly relevant given the current rapid developments of microfluidic devices \cite{Squires}.

Here we consider the behaviour of a liquid bridge formed between a pair of solid bodies. Investigations into such a liquid bridge geometry has a long history. Early studies of this problem have been performed, e.g. by Poisson \cite{Poisson}, Goldschmidt and Delauney \cite{Delauney} in the 19th century, who considered surfaces of revolution with constant mean curvature. More recent research in the field is motivated by its occurance in many different contexts. Examples include but are not limited to capillary condensation and evaporation \cite{Biben,Kohonen}, wet adhesion \cite{Vogel}, wet granular matter \cite{Herminghaus}, atomic force microscopy \cite{Weeks,AFM} and low gravity \cite{Langbein,Li} experiments. The existence of these liquid bridges often change the systems' physical properties because these bridges exert forces onto the surfaces, to which they are attached. 

Due to its broad relevance and implications, a number of aspects of the liquid bridge geometry have been addressed. The shape and force displacement relation of the liquid bridge have been analyzed for both flat \cite{Carter} and curved solid bodies \cite{Orr,Willett}. The stability of the liquid bridge shapes \cite{Langbein,Concus,Gillette} and the influence of gravity \cite{Langbein,Li,Orr} have also been investigated. These studies were primarily concerned with chemically homogeneous surfaces.

Recently, new patterning techniques have been developed that allow the construction of chemical surface domains on micrometer or even nanometer length scales. The corresponding liquid structures are small compared to the capillary length and gravity effects can be safely ignored. From the theoretical point of view, bridges between striped surface domains have been considered using both the theory of capillary surfaces as well as  lattice gas models \cite{Swain,Valencia}. In addition, polymeric nanobridges between two circular surface domains have also been studied by Molecular Dynamics simulations \cite{Yaneva}.

In this paper, we will extend these theoretical studies and consider liquid bridges in four different types of slab geometries or slit-pores: (i)  First, we will briefly discuss the case of two identical and chemically homogeneous surfaces, which are characterized by the same contact angle; (ii) Second, we will consider two chemically homogenous surfaces, which differ, however, in their contact angles, a geometry that has been previously studied in Ref. \cite{Souza1}; (iii) Next, one of the two surfaces is taken to contain a circular surface domain, the contact angle of which is smaller than the contact angle of the surrounding substrate surface; and (iv) Finally, both surfaces are decorated by circular surface domains.

In the first two cases, the two contact lines of the liquid bridge can move freely along the surfaces. In the cases (iii) and (iv), on the other hand, the contact lines may be pinned by the boundaries of the chemical surface domains. The pinning 'strength' depends on the wettability contrast between the chemical domains and the substrate surfaces. We will focus on a large wettability contrast, for which  the contact lines are essentially pinned to the domain boundaries. In the latter limit, case (iv) of the slab geometry or slit-pore as considered here leads to liquid bridges that are rather similar to those formed between two circular plates as studied e.g. in Ref. \cite{Gillette}, where the contact lines are pinned at the edges of the plates. 

In all four cases, we determine  the behavior of the liquid bridges close to their mechanical equilibrium states. We define mechanical equilibrium by those liquid bridge configurations that exert no force on the solid plates. We will analytically calculate the equilibrium distance of the plates and then, as they are slightly perturbed from equilibrium, we shall derive the effective spring constant of the liquid bridges. This is the first time, to the best of our knowledge, that the effective spring constant has been computed. To verify our analytical results, they will be compared with numerical calculations using the Surface Evolver package \cite{Brakke}.

We note that this is, in general, a very complicated problem because it depends on the details of the surface inhomogeneities which, on real surfaces, will be random in position and spans across multiple length scales. Thus, we shall limit ourselves to a subset of these possible scenarios. We focus on situations where the liquid bridge solutions still possess an axial symmetry. We shall assume that the solids are flat, smooth and parallel to each other, but the solid plates can be either chemically homogeneous or decorated by circular patches of more hydrophilic domains, compared to the remaining susbtrate surface.

This paper is organised as follows. In the next section, we will introduce our theoretical description as well as the methods that we will use to solve the problem. Then we present our main results: the equilibrium distance and the spring constant of the liquid bridge as a function of the volume, contact angle, and radius of the chemical domain. A conclusion summarises our results and compares them to relevant studies in the literature.

\section{Theoretical description}

\begin{figure}
\centering
\includegraphics[scale=1.0,angle=0]{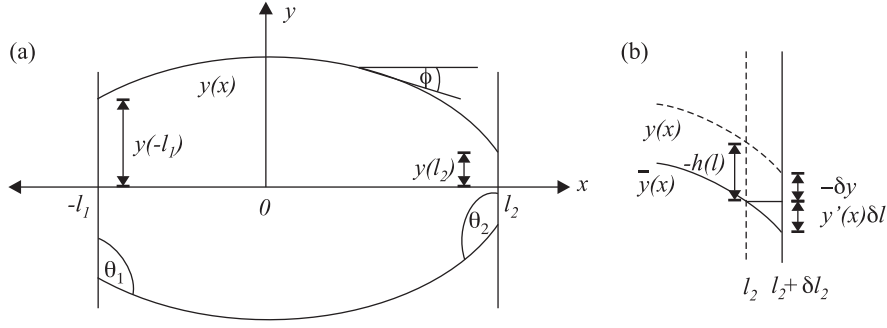}
\caption{(a) A schematic diagram of the liquid bridge geometry. The coordinate along the symmetry axis is denoted by $x$, the distance from this axis by $y$, and the local tilt angle by $\phi$. The liquid bridge contacts two parallel plates at $x=-l_1$ and $x=l_2$ with contact radii $y(-l_1)$ and $y(l_2)$, and contact angles $\theta_1$ and $\theta_2$ respectively. Note that in general $0\leq \theta_1,\theta_2\leq\pi$ and the contour $y(x)$ does not need to have an extremum. (b) Enlarged view close to the contact line as the right plate is displaced from $x = l_2$ (dashed line) to $l_2+\delta l_2$ (full line). The interface shape changes from $y(x)$ to $\bar{y}(x)$ and the contact line shifts by $\delta y$.}
\label{geometry}
\end{figure}

Consider the geometry shown in Fig. \ref{geometry}. A liquid $\beta$ is formed between two parallel plates $\sigma$ located at $x=-l_1$ and $x=l_2$ and is surrounded by a gas phase $\alpha$. Throughout this paper, we shall reserve the subscript 1 and 2 for the left and right walls respectively, as shown in Fig. \ref{geometry}. The plates are smooth, but they can either be chemically homogeneous or patterned with circular domains of radius $R$. In the latter case, let us denote the circular domain as $\gamma$ and the surrounding domain as $\delta$. We shall further assume that the contact angle of the circular domain $\theta_{\gamma}$ is smaller than the contact angle of the surroundings $\theta_{\delta}$. The contact angle is defined as the angle that the tangent of the liquid gas interface forms with the solid surface. For the geometry shown in Fig. \ref{geometry}, the contact angles $\theta_1$ and $\theta_2$ are related to the shape of the liquid gas interface $y(x)$ via
\begin{equation}
\cos{\theta_1} = - \frac{y'(-l_1)}{(1+y'(-l_1)^2)^{1/2}} 
\quad\quad \mathrm{and} \quad\quad 
\cos{\theta_2} = \frac{y'(l_2)}{(1+y'(l_2)^2)^{1/2}}, \label{CA}
\end{equation}
where the prime indicates a derivative with respect to $x$.

For a chemically homogeneous plate, the contact angle has a unique value given by the well-known Young's equation
\begin{equation}
\cos{\theta} = \frac{\Sigma_{\sigma\alpha}-\Sigma_{\sigma\beta}}{\Sigma_{\alpha\beta}}, \label{Young}
\end{equation}
where the $\Sigma$ symbols correspond to the interfacial tensions between the two phases denoted on the subscripts. The situation is more complex when the plate is chemically patterned. In this case, the value of the contact angle can vary between $\theta_{\gamma}$ and $\theta_{\delta}$ when the contact line is pinned at the domain boundary \cite{Lenz}. Inside and outside the circular domain, the contact angles are $\theta_{\gamma}$ and $\theta_{\delta}$ respectively.

If we restrict ourselves to axisymmetric solutions, the volume of the bridge is given by
\begin{equation}
V = \pi \int_{-l_1}^{l_2} y^2(x) dx \, .
\end{equation}  
The free energy $E$ of the bridge can be divided into two parts, a bulk term that accounts for the liquid-gas interfacial energy and two boundary terms which account for the walls. This decomposition has the form
\begin{eqnarray}
E &=& 2 \pi \Sigma_{\alpha\beta} \int_{-l_1}^{l_2} dx \left[y(1+y'^2)^{1/2} + \frac{\Delta P}{2\Sigma_{\alpha\beta}} y^2\right] \nonumber \\
&+& \sum_{l=-l_1,l_2} 2 \pi \int_0^{y(l)}y dy (\Sigma_{\sigma\beta}(y)-\Sigma_{\sigma\alpha}(y)) \, .
\end{eqnarray} 
As before, the $\Sigma$ symbols denote the interfacial tensions, and since we consider axisymmetric solutions, the variations of the interfacial tensions $\Sigma_{\sigma\beta}(y)$ and $\Sigma_{\sigma\alpha}(y)$ must be axisymmetric too. As shown in \cite{Lenz}, the contact angle equation (\ref{Young}) is valid for position dependent interfacial tensions as well. 

The pressure difference $\Delta P = P_{\alpha}-P_{\beta}$ is a Lagrange multiplier that we need because the volume of the liquid bridge is kept constant. As has been shown in the literature (e.g. \cite{Brinkmann}), the first variation of the free energy functional leads to two equations, an equation for the Laplace pressure
\begin{equation}
-\Delta P/\Sigma_{\alpha\beta} = 2M = \frac{1}{y(1+y'^2)^{1/2}}-\frac{y''}{(1+y'^2)^{3/2}}, \label{Euler} \\ 
\end{equation}
and the contact angle equation as given in eq \ref{Young}. In eq \ref{Euler}, $M$ is defined as the mean curvature of the liquid bridge. It is useful to note that for the parameterizations used here, the variations of $y(x)$ are automatically tangential to the walls. For general geometries, one has to be careful about the variations of the interface shape and of the contact line position.

Furthermore, using Noether's theorem, we find that the quantity 
\begin{equation}
f \equiv \frac{y}{(1+y'^2)^{1/2}} - My^2 \label{Noether}
\end{equation}
is a constant. In fact, as we shall see below, the constant $f$ is proportional to the force exerted by the liquid bridge onto the walls. 

To derive the force, suppose that we have a profile $y(x)$ that satisfies the Euler-Lagrange equation (eq \ref{Euler}) for a given drop volume and separation between the plates. If we now displace the plates from $[-l_1,l_2]$ to $[-l_1-\delta{}l_1,l_2+\delta{}l_2]$, the profile of the bridge changes slightly to $\bar{y}(x)=y(x) + h(x)$ and the energy of the system changes by
\begin{eqnarray}
\Delta E &=& 2 \pi \Sigma_{\alpha\beta} \int_{-l_1}^{l_2} dx \left[\frac{1+y'^2-yy''}{(1+y'^2)^{3/2}}-2My\right]h(x) \nonumber \\
&+& 2 \pi y(l_2)\left[\frac{\Sigma_{\alpha\beta}\,y'}{(1+y'^2)^{1/2}} +  (\Sigma_{\sigma\beta}-\Sigma_{\sigma\alpha}) \right]_{x = l_2}\delta{}y(l_2) \nonumber \\
&+&  2 \pi y(-l_1)\left[-\frac{\Sigma_{\alpha\beta}\,y'}{(1+y'^2)^{1/2}} +  (\Sigma_{\sigma\beta}-\Sigma_{\sigma\alpha}) \right]_{x = -l_1}\delta{}y(-l_1) \nonumber \\
&+& 2 \pi \Sigma_{\alpha\beta} \left[ \frac{y}{(1+y'^2)^{1/2}} -My^2\right]_{x = l_2}\delta{}l_2 \nonumber \\
&+& 2 \pi \Sigma_{\alpha\beta} \left[ \frac{y}{(1+y'^2)^{1/2}} -My^2\right]_{x = -l_1}\delta{}l_1 .
\end{eqnarray}
The first term is always zero because $y(x)$ is an extremal solution. The second and third terms can be zero in two cases, when the terms in the bracket go to zero or when $\delta{}y = 0$. The former is satisfied for a smooth homogeneous surface since they correspond to eq \ref{CA}, while the latter is true when the contact line is pinned at the boundary between $\gamma$ and $\delta$. In other words, for all the cases we shall consider here, only the fourth and fifth terms are important. With this in mind, the force can be written as
\begin{eqnarray}
F&=&-\frac{\delta{}E}{\delta{}(l_1+l_2)} = -2\pi\Sigma_{\alpha\beta}f \nonumber \\
&=& -2\pi\Sigma_{\alpha\beta}\left(y(l_2) \sin{\theta_2} - M y^2(l_2)\right) \label{force} \\
&=& -2\pi\Sigma_{\alpha\beta}\left(y(-l_1) \sin{\theta_1} - M y^2(-l_1)\right). \nonumber
\end{eqnarray}
The force $F$ as given by eq \ref{force} can be understood intuitively as the force acting on one wall and is equal to the superposition of the normal force component arising from the liquid-gas interface (first term) and the Laplace pressure of the liquid pushing on the wall (second term). When the force is positive (negative), the liquid bridge acts to increase (decrease) the separation between the two plates.

Let us now sketch how, in general, we can construct the shape of the liquid gas interface. We shall formulate the method in a similar way to Carter \cite{Carter}. Since the mean curvature $M$ and the force parameter $f$ are constants, we can parametrize the surface using the tilt angle $\phi = \phi(x) $ as defined in Fig. \ref{geometry}, which is directly related to the shape contour $y = y (x)$ via 
\begin{equation}
\tan \phi  = y'
\quad {\rm and} \quad 
\cos \phi = ( 1 + y'^2 )^{- 1/2}  \, . 
\end{equation}
Using the latter identity in eq \ref{Noether}, we obtain a quadratic equation for $y$ which has the two solutions
\begin{equation}
y_\pm = \frac{1}{2 M} ( \,  \cos \phi  \pm ( \cos^2 \phi  - 4 M f )^{1/2} \, ) \, .
\end{equation}
For small $f$, these two solutions approach the limiting values $y_{+} \approx \cos \phi / M$ and $y_{-} \approx 0$, respectively. Since a bridge should be characterized by  $y > 0$ for all $x$, we discard the solution $y_{-}$ and thus obtain the relation 
\begin{equation}
y = y_{+} = \frac{1}{2 M} ( \,  \cos \phi  + ( \cos^2 \phi  - 4 M f )^{1/2} \, ) \label{y-phi}
\end{equation}
between the shape contour $y$ and the tilt angle $\phi$. 

The $x$-coordinate can be obtained by using the fact that $y'(x) = \frac{dy}{d\phi}\frac{d\phi}{dx} = \tan{\phi}$. Solving for $\frac{dx}{d\phi}$ and integrating $\frac{dx}{d\phi}$ with respect to $\phi$ once, we obtain
\begin{equation}
x = -\frac{\sin{\phi}}{2M} - \frac{1}{2M} \int_0^{\phi} d\alpha \frac{\cos^2{\alpha}}{( \cos^2 \alpha  - 4 M f )^{1/2}}. \label{x-phi}
\end{equation}

Thus far, we have written our solutions in terms of two parameters, $f$ and $M$. The appropriate values for $f$ and $M$ come from our boundary conditions. At the walls, the two tilt angles
\begin{equation}
\phi(l_2) = \frac{\pi}{2} - \theta_2
\quad {\rm and} \quad
\phi(- l_1) = \theta_1 - \frac{\pi}{2} \, ,
\end{equation}
which further imply
\begin{equation}
\label{theta-phi}
\cos \theta_2 = \sin \phi_2
\quad {\rm and} \quad
\cos \theta_1 = - \sin \phi_1.
\end{equation}
These relations are direct consequences of  the  geometry in Fig. \ref{geometry}. Substituting these relations to eq \ref{x-phi}, we find that the distance between the two walls
\begin{eqnarray}
 L &=& l_1+l_2, \nonumber\\
 l_1 &=& -\frac{\cos{\theta_1}}{2M} - \frac{1}{2M} \int_0^{\pi/2-\theta_1} d\alpha \frac{\cos^2{\alpha}}{( \cos^2 \alpha  - 4 M f )^{1/2}}, \label{distance}\\
 l_2 &=& -\frac{\cos{\theta_2}}{2M} - \frac{1}{2M} \int_0^{\pi/2-\theta_2} d\alpha \frac{\cos^2{\alpha}}{( \cos^2 \alpha  - 4 M f )^{1/2}}.\nonumber
\end{eqnarray}

The second boundary condition comes from the conservation of volume
\begin{eqnarray}
V &=& \sum_{\theta=\theta_1,\theta_2} -\frac{\pi}{8M^3}\left[4\cos{\theta}-\frac{4}{3}\cos^3{\theta}-4Mf\cos{\theta}\right] \nonumber \\
&+& \sum_{\theta=\theta_1,\theta_2}   -\frac{\pi}{8M^3}\left[4 \int_0^{\pi/2-\theta} d\alpha \frac{\cos^4{\alpha}}{( \cos^2 \alpha  - 4 M f )^{1/2}}\right] \label{volume}\\
&+& \sum_{\theta=\theta_1,\theta_2}   -\frac{\pi}{8M^3}\left[-12Mf\int_0^{\pi/2-\theta} d\alpha \frac{\cos^2{\alpha}}{( \cos^2 \alpha  - 4 M f )^{1/2}}\right]. \nonumber
\end{eqnarray}

\section{Bridge morphologies for vanishing external forces}

The liquid bridge geometry encountered in many physical systems as well as its model experimental setups can be categorically divided into two arrangements. In the first category, the solid bodies are fixed and the liquid bridge arranges itself to minimize its interfacial energy. In the second category, at least one of the solid bodies is able to move and adjust its position. The theoretical formulation presented in the previous section is applicable for both categories, but from here onwards, we will focus on the latter.

In this section, we shall calculate the equilibrium distance of the solid surfaces when there is a liquid bridge connecting the two plates. We define the equilibrium distance as the distance at which the force exerted by the liquid bridge is zero.

When $F = f = 0$, the solution for the liquid gas interface simplifies considerably. In this case, eq \ref{y-phi} reads $y = \frac{1}{M} \cos \phi $ which implies
\begin{equation}
M y(l_2)  = \sin (\theta_2)
\quad {\rm and} \quad
M y(-l_1)  = \sin (\theta_1) \,  
\end{equation}
at the walls. Since a liquid bridge is characterized by $y(l_2)  > 0$ and $y(-l_1) > 0$, such a bridge must have positive mean curvature $M > 0$ for contact angles  $0 < \theta_1 < \pi$ and $0 < \theta_2 < \pi$. Furthermore, from eq \ref{distance}, we obtain the relation
\begin{equation}
\label{sum21}
\cos \theta_1 + \cos \theta_2  =  - M L < 0 ,
\end{equation}
which is equivalent to 
\begin{equation}
\label{sum2}
2 \cos \left(\frac{\theta_1+\theta_2}{2} \right) \, \cos \left(\frac{\theta_1- \theta_2}{2} \right) =  - M L < 0  \, . 
\end{equation}
Since $| \theta_1- \theta_2 | < \pi$ and  $ \cos \left(\frac{\theta_1- \theta_2}{2} \right) > 0$ for all possible values of the two contact angles,  it follows from eq \ref{sum2} that 
\begin{equation}
\theta_1+\theta_2 > \pi \, . \label{sum22}
\end{equation} 
Thus, a liquid bridge can only be in mechanical equilibrium for $F  = 0$ if the sum of the two contact angles exceeds 180$^\circ$. For $F=0$, the volume of the liquid bridge also has a relatively simple form
\begin{equation}
V = \pi{}l_1^3(\sec^2{\theta_1}-1/3) +  \pi{}l_2^3(\sec^2{\theta_2}-1/3).
\end{equation}

Let us now consider the solutions for some special cases. We can categorise them into symmetric (for identical walls) and asymmetric cases (for non identical walls). There are two possible scenarios for the former and three for the latter.

\subsection{Symmetric cases}
\begin{figure}
\centering
\includegraphics[scale=1.0,angle=0]{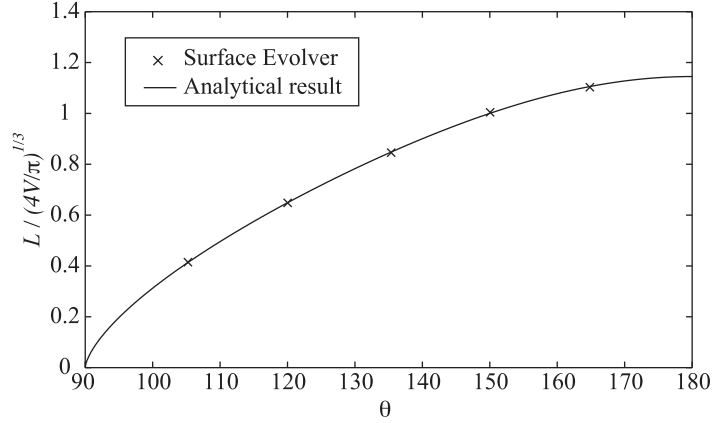}
\caption{Normalised plates separation $L/(4V/\pi)^{1/3}$ as a function of contact angle $\theta$ for mechanical equilibrium with external force $F=0$. Both solid surfaces are homogenous and have the same contact angle.}
\label{homo1}
\end{figure}

When the two walls are identical, 
\begin{eqnarray}
& y(-l_1) = y(l_2) = \sin{\theta}/M ,  \\
& l_1 = l_2 = L/2 = - \cos{\theta}/M , \\
& V = \frac{\pi}{4}L^3(\sec^2{\theta}-1/3).
\end{eqnarray}
The contact angle is well-defined for homogeneous plates and it is appropriate to characterise the equilibrium plate distance as a function of the liquid volume $V$ and the contact angle $\theta$. Fig. \ref{homo1} shows how the plates separation (normalised with respect to the liquid volume) depends on the contact angle. The normalised plates separation increases monotonically with respect to the contact angle.

\begin{figure}
\centering
\includegraphics[scale=1.0,angle=0]{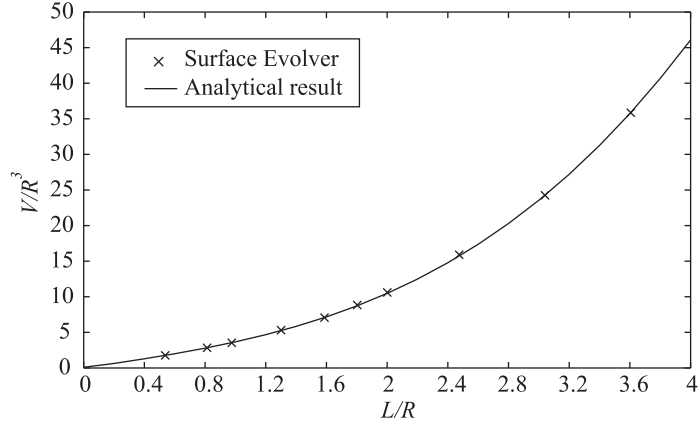}
\caption{Normalised liquid bridge volume $V/R^3$ as a function of normalised equilibrium distance $L/R$. Both solid surfaces are patterned with circular domains of radius $R$.}
\label{SE1}
\end{figure}

When the solid plates are patterned with circular patches, the contact line may be pinned at the boundary between domains $\gamma$ and $\delta$. In that case, the contact angle $\theta$ is not unique and its value can vary between $\theta_{\gamma}$ and $\theta_{\delta}$. On the other hand, the size of the circular patch $R$ sets an intrinsic length scale to the system. It is therefore more suitable to express the equilibrium distance as a function of $V$ and $R$. Substituting $y(-l_1) = y(l_2) = R$ and using the fact that $\sec^2{\theta} = 1 + \tan^2{\theta} = 1 + 4R^2/L^2$, we obtain
\begin{equation}
V = \frac{\pi}{4}L^3(2/3+4R^2/L^2) .
\end{equation}
The equilibrium contact angle itself may be expressed either as a function of the plate separation $L$ or the volume $V$
\begin{eqnarray}
&\tan{\theta} = -2R/L , \\
& V = -2\pi R^3 \cot^3{\theta} (2/3+\tan^2{\theta}). \label{V-teta-sym}
\end{eqnarray}
To illustrate how the liquid bridge volume depends on the equilibrium distance of the plates, this dependancy is plotted in Fig. \ref{SE1}. 

We have also checked our analytical result against Surface Evolver simulation results (data points in Figs. \ref{homo1} and \ref{SE1}) and it is clear that the analytical theory and numerical results compare favorably.

\subsection{Asymmetric cases} 
Our analytical relations can be used to study cases where the two walls are different, as long as the geometry remains axisymmetric. These include (I) two homogeneous plates of different contact angles, (II) one homogeneous and one chemically patterned plates, and (III) two chemically patterned plates. They are more realistic scenarios than the symmetric situations we have considered in the previous section. The solutions for all these three cases are given by:
\begin{eqnarray}
&(I)& V = \pi\left(\frac{\cos{\theta_1}}{\cos{\theta_2}}\right)^3l_2^3(\sec^2{\theta_1}-1/3) +  \pi{}l_2^3(\sec^2{\theta_2}-1/3), \\
&& L = l_2(1+\cos{\theta_1}/\cos{\theta_2}), \\
&(II)& V = -\pi{}l_2^3\cos^3{\theta_1} (1+R_2^2/l_2^2)^{3/2}(\sec^2{\theta_1}-1/3) +  \pi{}l_2^3(2/3+R_2^2/l_2^2), \\
&& L = l_2 (1 - \cos{\theta_1} (1+R_2^2/l_2^2)^{1/2}) , \\
&(III)& V = \pi\sqrt{R_2^2+l_2^2-R_1^2}\left(\frac{2}{3}l_2^2+\frac{2}{3}R_2^2+\frac{1}{3}R_1^2\right)+\pi{}l_2^3(2/3+R_2^2/l_2^2) , \\
&& L = l_2+\sqrt{R_2^2+l_2^2-R_1^2}.
\end{eqnarray}
Note that we have parameterised both the volume $V$ and the plates separation $L$ in terms of $l_2$, and for case (2), we have assumed that the right wall is chemically patterned with a patch of radius $R_2$ and the left wall is homogeneous.

\begin{figure}
\centering
\includegraphics[scale=1.0,angle=0]{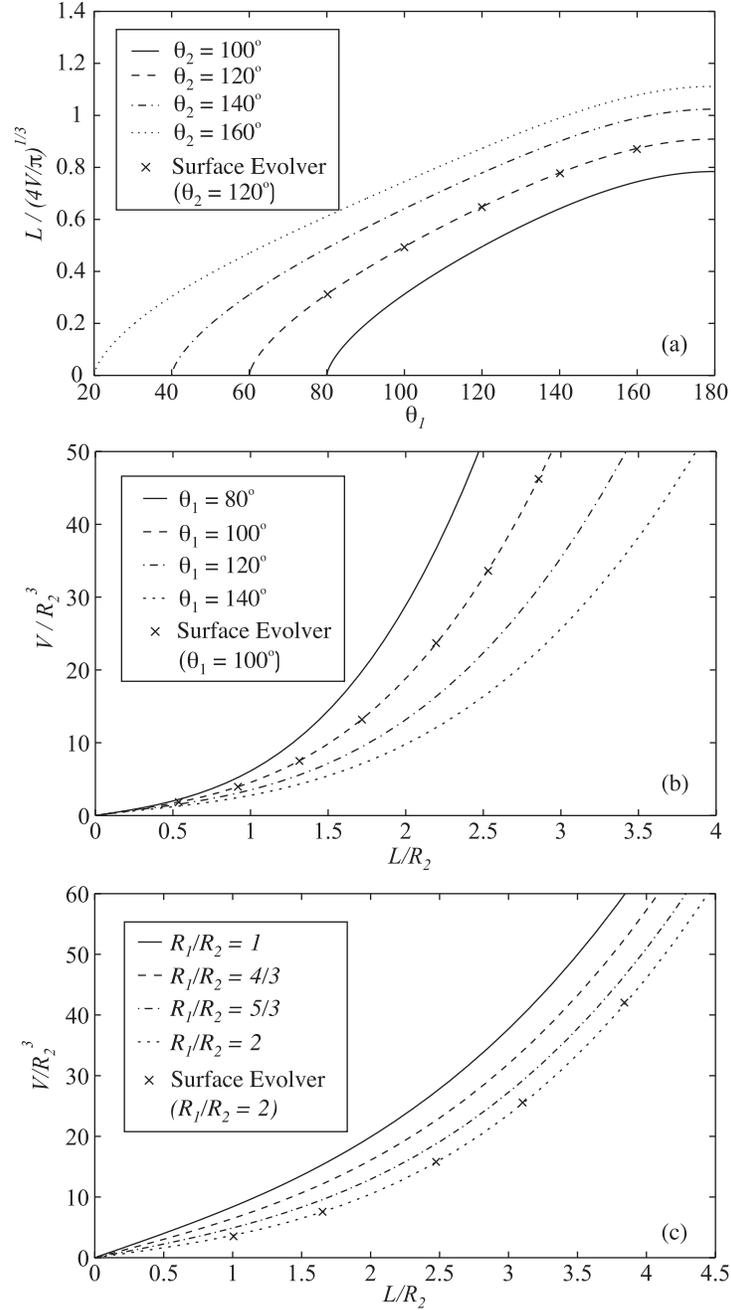}
\caption{Asymmetric plates. (a) Normalised plates separation $L/(4V/\pi)^{1/3}$ as a function of contact angle $\theta$ for two homogenous plates of different contact angles. (b-c) Normalised volume $V/R_2^3$ as a function of normalised plates separation $L/R_2$ for (b) one homogeneous and one chemically patterned plates, with domain radius $R_2$, and (c) two chemically patterned plates, with domain radii $R_1$ and $R_2$.}
\label{asymmetric}
\end{figure}

Fig. \ref{asymmetric}(a) shows how the normalised distance depends on the contact angles for case (I) and Fig. \ref{asymmetric}(b-c) how the normalised volume depends on the equilibrium plates separation for case (II) and (III) respectively. In Fig. \ref{asymmetric}(a-c), numerical results obtained using Surface Evolver are also shown for comparison for several representative cases. The qualitative understandings we obtain from the symmetric cases still apply in the asymmetric cases. In particular, the equilibrium distance and hence the liquid volume increases monotonically with the contact angle. One new and interesting aspect of the asymmetric cases is that for an equilibrium configuration to exist, one of the plate can be hydrophilic, or in the case of a patterned plate, one of the contact angles that the  liquid bridge forms can be less than $90^\circ$. This is in agreement with a recent study by De Souza {\it{et. al.}}\cite{Souza1}. However, we recall that for a liquid bridge to be stable, the sum of the contact angles $\theta_1+\theta_2 > 180^\circ$ (eq \ref{sum22}).

\section{Effective spring constants for small external forces}

When the plates are displaced from their equilibrium distance, there is a restoring force that pushes the system back towards its equilibrium state. This is true as long as one of the plates is free to move. One physical example is provided by an atomic force microscopy (AFM) experiment where the position of the AFM tip is not fixed \cite{Weeks,AFM}. In this case, one should in principle take into account that the AFM tip is curved, with a radius of curvature $R_{\mathrm{tip}}$. However, in the limit of $y(l) \ll R_{\mathrm{tip}}$, the curvature of the AFM tip can be neglected. The correction is of order $y(l)^2/R^2_{\mathrm{tip}}$.

In the previous section, we found the equilibrium distance for $F=f=0$. To obtain the spring constant of the liquid bridge, we use a dimensionless expansion parameter $\epsilon = Mf$. Subsituting this into eqs \ref{force}, \ref{y-phi}, \ref{distance}, and \ref{volume}, and keeping the first order correction terms in $\epsilon$, we obtain
\begin{eqnarray}
&\bar{M}\bar{F}=-2\pi\Sigma_{\alpha\beta} \, \, \epsilon ,\label{force1}\\
&\bar{M}\bar{y}(l)=\sin{\bar{\theta}}(1-\epsilon\csc^2{\bar{\theta}}) ,\label{width1}\\
&\bar{M}\bar{l}=-\cos{\bar{\theta}}-\epsilon\ln{(\csc{\bar{\theta}}+\cot{\bar{\theta}})} ,\label{distance1}\\
\mathrm{and} &-\frac{\bar{M}^3V}{\pi}= \sum_{\theta=\theta_1,\theta_2} \cos{\bar{\theta}}-\frac{1}{3}\cos^3{\bar{\theta}}-\epsilon\cos{\bar{\theta}} \label{volume1}.
\end{eqnarray}
The barred symbols correspond to the perturbed values of the parameters.

\subsection{Symmetric cases}

Let us first derive the spring constant for homogeneous and identical plates. We can use eqs \ref{distance1} and \ref{volume1} to obtain the relations
\begin{equation}
\Delta M = \frac{2\pi\cos{\theta}}{3VM^2} \epsilon, 
\end{equation}
\begin{equation}
\Delta l = \frac{-\Delta M l - \epsilon \ln(\csc{\theta}+\cot{\theta})}{M} ,
\end{equation}
for the change in curvature $\Delta M$ and displacement $\Delta l$ up to first order in $\epsilon$

Substituting the above equations into eq \ref{force1}, we find that the force-displacement relation is given by
\begin{equation}
F = -k \, 2\Delta l,
\end{equation}
with the spring constant
\begin{equation}
k \equiv \frac{-2\pi\Sigma_{\alpha\beta}}{2\left(\frac{\cos{\theta}}{2+\sin^2{\theta}}+\ln{(\csc{\theta}+\cot{\theta})}\right)}.
\end{equation}
The factor of 2 in front of $\Delta l$ comes from our convention to define the distance between the plates as $2l$ for the symmetric cases.

The derivation for the patterned plates is slightly more elaborate. First, we need to solve eqs \ref{width1} and \ref{volume1} simultaneously to obtain the relations
\begin{eqnarray}
(1+\cos^2{\theta}) \Delta\theta  = 2\epsilon\cot{\theta} ,  \\
\Delta M R = \cos{\theta} \Delta\theta - \epsilon\csc{\theta} ,
\end{eqnarray}
for the change in curvature $\Delta M$ and contact angle $\Delta \theta$ up to first order in $\epsilon$. Then we substitute our results for $\Delta M$ and $\Delta\theta$ into eq \ref{distance1} to obtain the displacement $\Delta l$
\begin{equation}
\Delta l = \frac{-\Delta M l + \sin{\theta}\Delta\theta - \epsilon \ln(\csc{\theta}+\cot{\theta})}{M} .
\end{equation}
Finally, after some algebra, we find that the spring constant $k$ in the force-displacement relation is given by
\begin{equation}
k = \frac{-2\pi\Sigma_{\alpha\beta}}{2\left(-\frac{\cos{\theta}}{1+\cos^2{\theta}}+\ln{(\csc{\theta}+\cot{\theta})}\right)}. \nonumber
\end{equation}

\begin{figure}
\centering
\includegraphics[scale=1.0,angle=0]{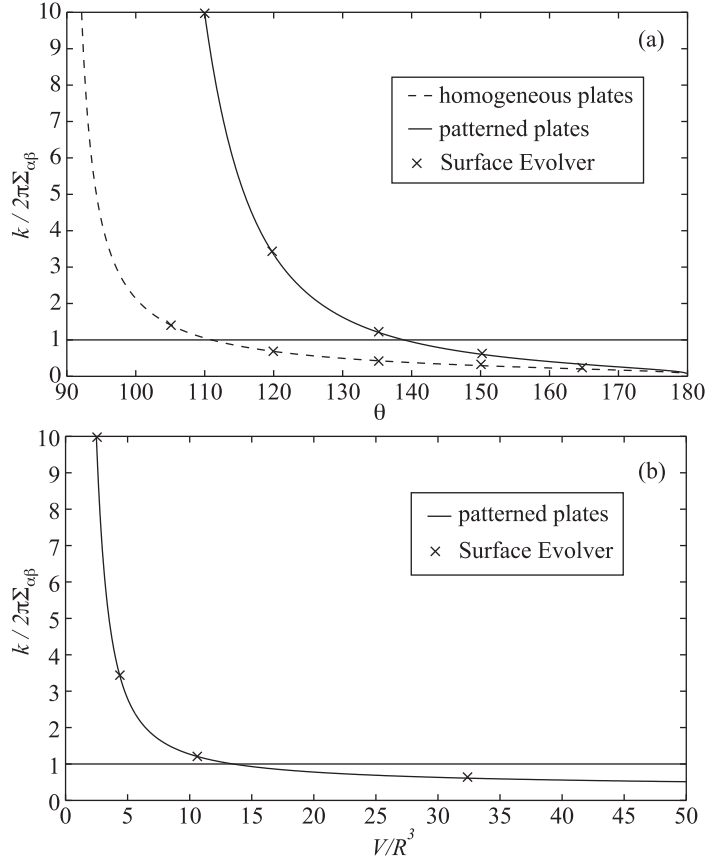}
\caption{Symmetric plates. Normalised spring constant of the liquid bridge $k/2\pi\Sigma_{\alpha\beta}$ is plotted as a function of (a) equilibrium contact angle $\theta$ for both the smooth and patterned surfaces scenarios and (b) normalised liquid volume $V/R^3$ for the patterned surfaces case.}
\label{spring}
\end{figure}

We plot in Fig. \ref{spring}(a) how the spring constant of the liquid bridge depends on the equilibrium contact angle for both the homogeneous and patterned surfaces. For a given contact angle, the spring constant is stiffer for the patterned surfaces than for the homogeneous surfaces. The spring constant also diverges at $\theta=90^\circ$, as $k\propto\frac{1}{(\theta-\pi/2)}$ and $k\propto\frac{1}{(\theta-\pi/2)^3}$ for the homogeneous and patterned plates respectively. Then it decreases monotonically with respect to the contact angle. This is what one would expect intuitively. The larger the contact angle, the weaker the liquid bridge is bounded to the surface and hence it is easier to displace them from equilibrium. Furthermore, as we shall see below for the asymmetric cases, a generic feature of our calculations is that the spring constant diverges whenever $\theta_1+\theta_2=180^\circ$. Naturally, for two symmetric plates this occurs at $\theta=90^\circ$.

For patterned walls, the liquid bridge is better parameterised as a function of volume rather than contact angle. The spring constant is plotted in Fig. \ref{spring}(b) as a function of the normalised volume $V/R^3$. We have used eq \ref{V-teta-sym} to relate the liquid bridge volume and the contact angle. In terms of the liquid bridge volume, the divergence happens when $V/R^3$ approches zero.

As in the previous section, we also compare our analytical results against Surface Evolver simulations. The latter are shown as data points in Fig. \ref{spring} and they agree well with the analytical formulas.

\subsection{Asymmetric cases}

The same machinery can be applied to cases where the two plates are not identical. One needs to expand $\bar{y}$, $\bar{l}$, $\bar{M}$, and $\bar{\theta}$ to first order in $\epsilon$ and then solve eqs \ref{force1}-\ref{volume1} self consistently. Below we write down the results for the asymmetric cases. As in the previous section, we consider (I) two homogeneous plates of different contact angles, (II) one homogeneous and one chemically patterned plates, and (III) two chemically patterned plates.
\begin{eqnarray}
&(I)\,\,(-k/2\pi\Sigma_{\alpha\beta})^{-1} = &\frac{(\cos{\theta_1}+\cos{\theta_2})^2}{3\cos{\theta_1}-\cos^3{\theta_1}+3\cos{\theta_2}-\cos^3{\theta_2}} + \\
&&\ln{(\csc{\theta_1}+\cot{\theta_1})} + \ln{(\csc{\theta_2}+\cot{\theta_2})} , \nonumber \\
&(II)\,\,(-k/2\pi\Sigma_{\alpha\beta})^{-1} = &\frac{(\cos{\theta_1}+\cos{\theta_2})\cos{\theta_1}}{3\cos{\theta_1}-\cos^3{\theta_1}+3\cos{\theta_2}-\cos^3{\theta_2}} + \frac{\cos{\theta_2}}{\sin^2{\theta_2}} + \\
&&\ln{(\csc{\theta_1}+\cot{\theta_1})} + \ln{(\csc{\theta_2}+\cot{\theta_2})} - \nonumber \\
&&\left(\frac{2}{\sin{\theta_2}}-\frac{2\sin^3{\theta_2}\cos{\theta_1}}{3\cos{\theta_1}-\cos^3{\theta_1}+3\cos{\theta_2}-\cos^3{\theta_2}}\right) \times \nonumber \\
&&\left(\frac{2\cot{\theta_2}+2\cos{\theta_1}/\sin{\theta_2}+\sin^2{\theta_1}\cos{\theta_1}/\sin{\theta_2}-\sin{\theta_2}\cos{\theta_1}}{2+2\cos^2{\theta_2}+4\cos{\theta_2}\cos{\theta_1}+2\sin^2{\theta_1}\cos{\theta_1}\cos{\theta_2}}\right), \nonumber \\
&(III)\,\,(-k/2\pi\Sigma_{\alpha\beta})^{-1} = &-\frac{2}{\sin{\theta_1}}f(\theta_1,\theta_2)-\frac{2}{\sin{\theta_2}}f(\theta_2,\theta_1) +\frac{\cos{\theta_1}}{\sin^2{\theta_1}}+\frac{\cos{\theta_2}}{\sin^2{\theta_2}}+\\
&&\ln{(\csc{\theta_1}+\cot{\theta_1})} + \ln{(\csc{\theta_2}+\cot{\theta_2})} ,\nonumber
\end{eqnarray}
where in the last equation $f(\alpha,\beta)$ is defined as
\begin{displaymath}
f(\alpha,\beta) = \frac{2\cot{\alpha}+2\frac{\cos{\beta}}{\sin{\alpha}}+\frac{\sin^2{\beta}\cos{\beta}}{\sin{\alpha}}-\sin{\alpha}\cos{\beta}+2\sin^3{\beta}\sin{\alpha}\,\frac{R_{\beta}\csc{\alpha}-R_{\alpha}\csc{\beta}}{2R_{\alpha}\cos{\beta}}}{2+2\cos^2{\alpha}+4\cos{\alpha}\cos{\beta}+2\sin^2{\beta}\cos{\beta}\cos{\alpha} +2\sin^3{\beta}\sin{\alpha}\,\frac{R_\beta\cos{\alpha}}{R_\alpha\cos{\beta}}} .\nonumber
\end{displaymath}
and the two contact angles are related via $R_1/R_2=\sin{\theta_1}/\sin{\theta_2}$.
\begin{figure}
\centering
\includegraphics[scale=1.0,angle=0]{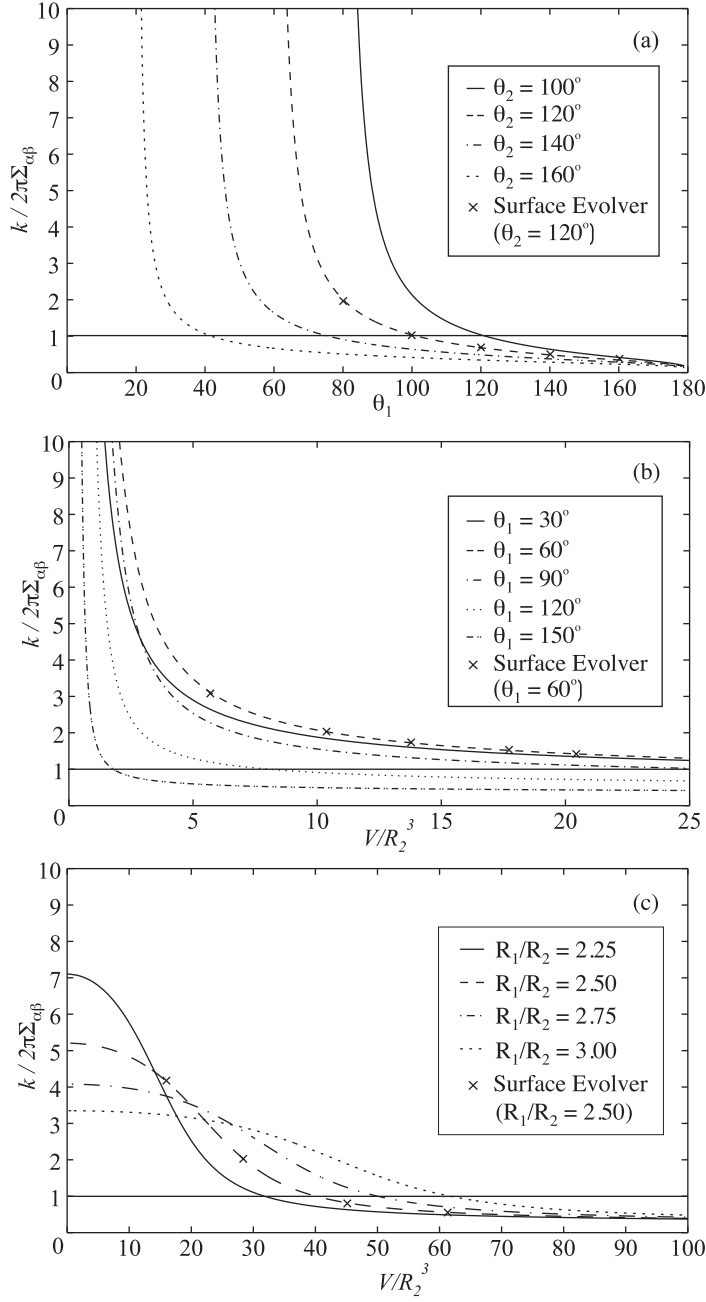}
\caption{Asymmetric plates. Normalised spring constant of the liquid bridge $k/2\pi\Sigma_{\alpha\beta}$ is plotted as a function of (a) equilibrium contact angle $\theta$ for two different homogeneous plates and (b-c) normalised liquid volume $V/R_2^3$ for (b) one homogeneous and one chemically patterned (domain radius $R_2$) plates and (c) two chemically patterned plates with different radii of chemical domains $R_1$ and $R_2$.}
\label{springa}
\end{figure}

The spring constants are plotted as functions of the relevant variables in Fig. \ref{springa} for the asymmetric cases, with the data points corresponding to Surface Evolver calculations for several representative cases. In Fig. \ref{springa}(a), it is plotted as a function of the contact angles of the two homogenous plates. As we have mentioned in the previous section, for asymmetric plates, equilibrium liquid bridge configurations do not require both walls to be hydrophobic. The effective spring constant can therefore be defined as long as $\theta_1+\theta_2 > 180^\circ$. When $\theta_1+\theta_2 = 180^\circ$, the spring constant diverges and as before, its value decreases monotonically with increasing $\theta_1$ and $\theta_2$. It can be shown that the divergence goes as $k\propto\frac{1}{(\theta_1+\theta_2-\pi)}$.

The divergence in the spring constant is also observed for case (II), again when $\theta_1+\theta_2 = 180^\circ$ or in terms of volume when the liquid bridge volume approaches zero. This is shown in Fig. \ref{springa}(b). In addition, this scenario shows one interesting behaviour. For a given liquid bridge volume, there is an optimal contact angle (for the homogeneous plate) at which the spring constant is maximum. This is due to the fact that the spring constant depends on the contact angles on both surfaces. When $\theta_1$ is large, the spring constant is relatively weak as expected. When $\theta_1$ is small, the equilibrium contact angle at the patterned wall $\theta_2$ is large and as a result the spring constant is again relatively weak. The spring constant is relatively stiff when the values of the two contact angles are moderate.

Finally, the spring constant is shown as a function of the liquid bridge volume and the ratio of the radii of the chemical patches for case (III) in Fig. \ref{springa}(c). Unlike the two asymmetric cases we have considered above, we observe no divergence in this scenario. Instead, the spring constant reaches a plateau as the liquid bridge volume approaches zero. We previously note that the divergence is observed whenever $\theta_1+\theta_2=180^\circ$. Here the two contact angles are related via $\sin{\theta_1}/\sin{\theta_2} = R_1/R_2$ and as a result, the condition $\theta_1+\theta_2=180^\circ$ is never satisfied for $R_1 \neq R_2$. The maximum value of spring constant and the rate at which its value decreases as a function of volume depends strongly on the ratio of the radii of the chemical domains. When the ratio approaches 1, the maximum spring constant is very large and it decays rapidly with volume. This is expected since we ought to recover the symmetric scenario when the ratio is 1. On the other hand, when the ratio is large, the maximum spring constant is small and it decays very slowly with volume. To be more precise, here we have set $R_2$ as the typical length scale and the variation is slow relative to this length scale.

\section{Summary}

To conclude, we have theoretically studied the behaviour of a liquid bridge formed between a pair of rigid and parallel plates at and close to mechanical equilibrium, corresponding to vanishing and small external force $F$. We have computed the equilibrium distance of the liquid bridge and derived its effective spring constant as it is perturbed from equilibrium. The analytical results are then verified against Surface Evolver simulations.

Our results for the equilibrium distance is identical to Carter's \cite{Carter} for the identically homogeneous plates scenario and compare very well with recent molecular dynamics simulations by Yaneva et. al. \cite{Yaneva} for surfaces patterned with circular domains of radius $R$. In their paper, Yaneva et. al. tentatively used the relation $\tan{\theta} = -2R/L$ derived by Swain and Lipowsky \cite{Swain} to fit their data for the equilibrium contact angle. Swain and Lipowsky derived this relation when they considered a liquid bridge formed between two parallel stripes. This is essentially the two dimensional version of the problem considered in this paper. Our results here show that the full three dimensional treatment of the problem gives an identical relation and hence explain why the molecular dynamics simulations by Yaneva et. al. fit the theory by Swain and Lipowsky very well even though they were done for different geometries. We have also extended our calculations to include cases where the two plates are not symmetric. These cases are more realistic and relevant to real experimental systems. Whereas in the symmetric cases the plates have to be hydrophobic for the liquid bridge to be in mechanical equilibrium with $F=0$, this condition is relaxed for the asymmetric scenarios as long as the sum of the contact angles is larger than $180^\circ$, see eq \ref{sum2}.

Our model predicts that surface patterning plays an important role in the effective stiffness of the liquid bridge. We find that the spring constant diverges whenever the sum of the contact angles is $180^\circ$, see Figs. \ref{spring}(a) and \ref{springa}(a). This can be achieved for all the cases we have considered here, except when the two walls are patterned with different radii of chemical domains. In the later case, $\theta_1+\theta_2 > 180^\circ$ and the spring constant is always finite. Another generic property we observe is that the spring constant gets softer with increasing contact angle and volume. However, the maximum value of the spring constant and the rate at which it decreases depends on the properties of the two plates. For example, it is possible to devise a configuration where the spring constant is relatively soft for a large range of parameters. 

In our calculations, we have assumed that when the plates are patterned, the contact line is pinned at the boundary between the chemical domains, $\gamma$ and $\delta$. Such contact line pinning can also be provided by topographical patterns, e.g. by circular grooves or posts. Furthermore, the contact line may also depin as one varies an appropriate control parameter, such as the liquid volume. In the case of chemical domains, the contact line will move outward when $\theta > \theta_{\delta}$ and inward when  $\theta < \theta_{\gamma}$. In this scenario, the results we have derived here for the patterned plates are only valid for contact angles $\theta$ that satisfy $\theta_{\delta} > \theta > \theta_{\gamma}$. When the contact line is completely in the $\gamma$ and $\delta$ domains, one can simply use the results for the homogeneous plates. One may also need to take into account the intrinsic contact angle hysteresis of the surfaces \cite{Souza2}. Here we define intrinsic contact angle hysteresis as the variation in the value of the contact angle due to roughness or chemical heterogeneities at length scale which is much smaller than $R$ and $V^{1/3}$. In that case, depending on whether the contact line is advancing or receding, one should appropriately substitute the contact angle in our formula with the advancing or receding contact angle.
 
Another potential advantage of chemical patterning is that it may offer a simple way to stabilize liquid bridges against tilt perturbations. For two parallel homogeneous plates, a slight tilt or misalignment can result in a substantial displacement of the liquid bridge \cite{Concus}. Such structural instability is generally an unwanted feature, although in some special cases, such as the transportation of water droplets by shorebirds \cite{Prakash}, it can be of use. While it remains to be shown by explicit calculations, we expect that contact line pinning at the boundaries between the chemical domains will hinder the displacement of the liquid bridge, at least for moderate tilt angles.

The results presented in this paper are relevant for a number of different physical systems where the liquid bridge geometry plays a role. For examples, Vogel and Steen \cite{Vogel} have recently proposed a switchable capillary based adhesion device, and our calculations suggest a number of ways in which surface patterning can be used to control the stiffness of the capillary bridges and hence the strength of the adhesion. In wet granular system \cite{Herminghaus}, liquid bridges form and they strongly influence the stability of the system. Here the curvatures of the beads may become important. Thus, one useful extension of the current work will be to investigate such effects on the equilibrium distance and spring constant of the liquid bridges, particularly in situations when they are not symmetric. It is also of great interests to investigate the dynamical response of the above mentioned physical systems near equilibrium. For such studies, in addition to the results presented here, one has to take into account the viscosity of the liquid and the friction of the liquid bridge at the parallel planes, as well as solve its hydrodynamic equations of motion.

\end{document}